\documentclass[conference]{IEEEtran}
\IEEEoverridecommandlockouts
\usepackage{cite}

\usepackage[left=.6in,right=.6in,top=.6in,bottom=.9in]{geometry}
\usepackage{amsmath,amssymb,amsfonts}
\usepackage{algorithmic}
\usepackage{graphicx}
\usepackage{textcomp}
\usepackage{xcolor}
\PassOptionsToPackage{hyphens}{url}\usepackage{hyperref}
\usepackage{svg}
\usepackage{setspace}
\usepackage{fancyhdr}
\def\BibTeX{{\rm B\kern-.05em{\sc i\kern-.025em b}\kern-.08em
    T\kern-.1667em\lower.7ex\hbox{E}\kern-.125emX}}
\begin{document}

\title{Deepfakes, Misinformation, and Disinformation in the Era of Frontier AI, Generative AI, and Large AI Models}

\author{\IEEEauthorblockN{Mohamed R. Shoaib, Zefan Wang, Milad Taleby Ahvanooey, Jun Zhao}
\IEEEauthorblockA{{School of Computer Science and Engineering} \\
{Nanyang Technological University}\\
Singapore\\
All authors contribute equally to the paper.\\ Mohamedr003@e.ntu.edu.sg, Zefan.wang@ntu.edu.sg, Milad.ta@ntu.edu.sg, Junzhao@ntu.edu.sg}}

\maketitle
\thispagestyle{fancy}
\pagestyle{fancy}
\chead{This paper appears in IEEE International Conference on Computer Applications (ICCA), 2023.}
\cfoot{\thepage}
\renewcommand{\headrulewidth}{0.4pt}
\renewcommand{\footrulewidth}{0pt}

\begin{abstract}
With the advent of sophisticated artificial intelligence (AI) technologies, the proliferation of deepfakes and the spread of m/disinformation have emerged as formidable threats to the integrity of information ecosystems worldwide. This paper provides an overview of the current literature. Within the frontier AI's crucial application in developing defense mechanisms for detecting deepfakes, we highlight the mechanisms through which {generative AI based on large models (LM-based GenAI) craft seemingly} convincing yet {fabricated} contents. We explore the multifaceted implications of LM-based GenAI on society, politics, and individual privacy {violations}, underscoring the urgent need for robust defense strategies. To address these challenges, {in this study, we} introduce an integrated framework that combines advanced detection algorithms, cross-platform collaboration, and policy-driven initiatives to mitigate the risks associated with AI-Generated Content (AIGC). By leveraging multi-modal analysis, digital watermarking, and machine learning-based authentication techniques, we propose a defense mechanism {adaptable to AI capabilities of ever-evolving nature}. Furthermore, the paper advocates for a global consensus on the ethical usage of {GenAI and implementing cyber-wellness educational} programs to enhance public awareness and resilience against m/disinformation. Our findings suggest that a proactive and collaborative approach involving technological innovation and regulatory oversight is {essential for safeguarding netizens while interacting with cyberspace against the insidious effects of deepfakes and GenAI}-enabled m/disinformation campaigns.

\end{abstract}

 
\section{Introduction}

The frontier AI, characterized by its advanced capabilities and cutting-edge applications, significantly enhances the realism of deepfakes \cite{anderljung2023frontier}. Concurrently, it is instrumental in devising innovative solutions to detect and counter m/disinformation. Frontier AI encompasses new, innovative AI technologies that could exhibit sufficiently dangerous capabilities such as generative AI, advanced machine learning algorithms, large models, etc. The implications of frontier AI technologies extend beyond technological advancements, necessitating a global consensus on ethical tools usage and the implementation of comprehensive cyber-wellness educational programs\footnote{{https://www.whitehouse.gov/briefing-room/statements-releases/2023/10/30/fact-sheet-president-biden-issues-executive-order-on-safe-secure-and-trustworthy-artificial-intelligence/}}. Such measures are critical in equipping society to navigate the complex dynamics of information dissemination and integrity in the era of frontier AI\footnote{{https://www.channelnewsasia.com/singapore/ai-safety-summit-singapore-pm-lee-frontier-3892476}}.

{Over the last decade, the precipitous advancements in Generative Artificial Intelligence with large models (LM-based GenAI) have made revolutionary progress in crafting human-like multimedia content (e.g., text, image, video, or audio). Foundation models are a form of LM-based adaptable models that have become the backbone of significant} technological progress, driving innovations from autonomous vehicles to personalized medicine \cite{cao2023comprehensive}. However, {considering the power of LM-based GenAI tools, they might bring unprecedented risks and unintended consequences to our society, for instance, by empowering malicious actors to apply for cyber-scamming or cyberbullying in the form of deepfake advertisements through social media platforms \cite{mirsky2021creation}. This paper delves into these phenomena by discussing both possible outcomes of LM-based GenAI models, their societal impacts, and the urgent need of today's society for comprehensive defense mechanisms and sufficient cyber-wellness programs \cite{shin2022parental}}.

The rise of \textbf{\textit{Deepfake}}, a portmanteau of ``deep learning" and ``fake" media, are digital fabrications {in which realistic likenesses of things are synthetically generated or entirely altered to say or do something that never occurred \cite{danry2022ai}. Due to the public accessibility of sophisticated LM-based GenAI tools (e.g., ChatGPT and LivePerson), anyone can craft deepfake contents.} As these capabilities become democratized, the potential for misuse scales exponentially. Mis/disinformation, closely related but not limited to deepfakes, encompasses all forms of false or misleading information deliberately spread to deceive {netizens (active Internet users)}. This phenomenon is not new; however, the advent of LM-based GenAI models has supercharged its potential reach and believability. {Multimedia contents (e.g., texts, images, audios, and videos) produced by the LM-based GenAI tools can now fabricate reality so that discovering truth from fiction becomes increasingly challenging (e.g., family voice cloning threats)~\cite{amezaga2022availability}.}

The implications of the LM-based GenAI technologies are profound and multifaceted. Democracies worldwide grapple with the ramifications of AI generated content (AIGC) on electoral processes and public opinion \cite{danry2022ai}. {Netizens} face unprecedented threats to their privacy and security, as the {deepfakes that might be created of their public data} may act without their consent or even knowledge. Furthermore, the media landscape, the traditional {outlets} of factual information {propagation}, is undergoing a seismic shift as journalists and content creators confront the existential question of {what is a piece of trustworthy information in the post-deepfake era \cite{fletcher2018deepfakes}}.\vspace{-3pt}

\section{Background\vspace{-1pt}}

 This section gives the background for our discussion.\vspace{-2pt} 

\subsection{Historical Context of Information Manipulation\vspace{-2pt} }

Historically, information manipulation was labor-intensive and required significant resources, restricting its practice to powerful entities such as state {officials} or large organizations~\cite{zmud1990opportunities}. The infamous propaganda of wartime misinformation campaigns, psychological operations, and political machinations are {some testaments of how entities will affect} public opinion or discredit opposition~\cite{silverman2021seeing}. The advent of digital technology began a shift, enabling broader participation in information manipulation with the rise of Photoshop, video editing, {social media platforms, and LM-based GenAI tools} to disseminate such content widely and rapidly \cite{ahvanooey2020anitw}.

\subsection{Frontier AI Amplifying and Combating Digital Deception}
Frontier AI has reshaped the challenges in information manipulation. Its advances in neural networks and machine learning have heightened deepfakes realism, complicating the distinction between real and fake content\footnote{{https://www.gov.uk/government/publications/frontier-ai-taskforce-first-progress-report/frontier-ai-taskforce-first-progress-report}}. Concurrently, frontier AI is crucial in developing tools to counter misinformation and disinformation, as highlighted in recent studies. This dual role underscores both its potential for generating and detecting digital falsehoods.

\subsection{Evolution of {LM-based GenAI Tools} in Media Creation}

The role of {LM-based GenAI tools} in media creation started benignly enough, with techniques designed to enhance image quality, recommend content, or power voice assistants. As machine learning models advanced, they transcended these supportive roles, {becoming regular tools} in content creation. Generative adversarial networks (GANs)~\cite{goodfellow2014generative}, introduced in 2014, represented a significant leap forward, enabling the creation of photorealistic images indistinguishable from {actual photographs by the unaided vision systems. The evolution of AI continues with LM-based GenAI} that could synthesize human voices, compose music, and create realistic video footage \cite{cao2023comprehensive}.

%I suggest we eliminate the following subsection as its almost the same content as what we have described in the introduction.

\subsection{AI-Generated Mis/Disinformation}

{Technically, if deepfakes are generated based on event-related concepts, they could be formed as} mis/disinformation~\cite{zhou2023synthetic}. While {text manipulation is less technologically complex than other media files}, the implications are no less severe. Automated {ChatBots can disseminate false information by deploying LM-based GenAI tools that can craft fake news articles by claiming to be written by reputable sources. Malicious actors can deploy these ChatBots to spread using social media platforms, which} can inadvertently prioritize and amplify misleading content.

\subsection{Previous Efforts in Combating Digital Misinformation}

{{In the literature, many researchers have taken promising steps to} counter digital misinformation involving content moderation, community reporting, and algorithmic detection \cite{cao2023comprehensive, mirsky2021creation}. However, these methods face challenges, such as the overwhelming content volume and evolving misinformation techniques. LM-based GenAI models play a significant role in spreading {deepfakes that may cause mis/disinformation,} necessitating a deeper understanding of effective defense strategies \cite{he2023survey, siwakoti2023less}.}

\section{The Rise of Large AI Models}

The third decade of the 21st century {is considered the landmark of} a turning point in the capabilities of artificial intelligence, primarily through the advent of {LM-based GenAI}. AI foundation models and LM-based GenAI models (i.e., LLMs, LVMs, LAMs, or LMMs) have demonstrated unprecedented proficiency in understanding and generating human-like text, images, and sounds, leading to significant advancements in AIGC{\cite{hadi2023large}}. This section outlines the development of LM-based GenAI models, their capabilities, and {their associated risks}.

\subsection{Overview of LM-based GenAI Models}

LM-based GenAI models, such as OpenAI's GPT (Generative Pre-trained Transformer) series~\cite{brown2020language}, Google's BERT (Bidirectional Encoder Representations from Transformers)~\cite{devlin2018bert}, and others, represent the {modern cutting edge technology, which craft contents automatically. These models include Large Language Models (LLMs), Large Vision Models (LVMs), Large Audio Models (LAMs), or Large Multimodel Models (LMMs)} are characterized by their deep neural networks, which consist of millions{ or even billions} of parameters that enable them to process and generate complex data patterns. The ``large" in their name not only denotes their size in terms of parameters but also their vast training datasets and substantial computational power required for {performing} their operations.

\subsection{Training and Functioning}

The training process of {the LM-based GenAI} models involves feeding them enormous datasets, often sourced from the Internet, including books, articles, websites, and other {publicly available} media. This training allows the LM-based GenAI models to learn the nuances of human language, visual cues, and audio patterns {\cite{biderman2023pythia}}. They function by predicting the next word in a sentence, the next pixel in an image, or the {following} waveform in an audio file, learning from context and mimicking the style and {texture} of their training data {\cite{ xiao2023smoothquant}}.

%in my opinion, the following section is somehow redundant as its content was mentioned in the above sections many times.

\subsection{Case Studies of Deepfakes and {their Associated M/disi}nformation}

Real-world instances of {misusing LM-based GenAI} technologies provide sobering case studies. Deepfake videos have been used to create fake celebrity {advertisements,} pornographic videos, fabricated political speeches, and {voice cloning of} CEOs to commit fraud. AIGC has been employed to create fake news articles and social media posts that have gone viral, influencing public opinion and potentially affecting election outcomes{\cite{xu2023combating}}.

\subsection{Risks Associated with LM-based GenAI Capabilities}
{Evidently, the risks these models pose are beyond their capabilities as they provide opportunities for academic misconduct \cite{bin2023use} or deepfake phishing \cite{mirsky2023threat} and many uncovered threats. The fact that LM-based GenAI tools can be deployed in deceptive scenarios as convincing m/disinformation provides opportunities} for virtually anyone with the requisite technical know-how to launch sophisticated {misleading} campaigns. The potential for these technologies to be used for blackmail, electoral interference, and social unrest is a pressing concern{\cite{mustak2023deepfakes}}. Moreover, the speed at which AIGC can be produced outstrips the ability of current detection and moderation systems to keep up, creating a game of digital cat-and-mouse where the mouse is increasingly agile.

\section{Societal Implications}

The societal implications of deepfakes and { mis/disinformation generated by LM-based GenAI are bringing unprecedented impacts}, touching upon every facet of modern life—from politics and security to individual rights and societal trust{\cite{gregory2023fortify}}. {In the following, we provide an overview of} the far-reaching consequences of these phenomena and underscore the critical need for a robust societal response.

\subsection{Effects on Democracy and Public Opinion}

In democratic societies, the integrity of public discourse is foundational. Deepfakes {can be deployed as LM-based GenAI-generated misinformation that threaten the integrity of news propagation, as they could be exploited for fabricating scandals, falsifying} records of public statements, and manipulating electoral processes. When voters cannot distinguish between real and {falsified} representations of candidates or policies, the very fabric of democratic decision-making is undermined{\cite{schiff2023liar}}. The dissemination of spurious information can sway elections, fuel political polarization, and erode the public's trust in democratic {governments}.

\subsection{Impact on Privacy and Personal Security}

The ability to create convincing fake images and videos of individuals without {getting their} consent has raised alarm bells regarding privacy and personal security. Deepfakes can be weaponized to discredit individuals, exploit them for blackmail, or invade their privacy in egregious ways, as seen in the creation of non-consensual deepfake pornography{\cite{ciftci2023my}}. {The impacts of such artcrafts are deceptive effects} on free expression and the pervasive sense of vulnerability as individuals grapple with the potential for their likeness to be used in harmful ways.

\subsection{Consequences for Media and Journalism}

{Technically,} Journalism's role as the fourth estate is predicated on the ability to provide accurate, reliable information. Deepfakes and AIGC pose existential challenges to this role. Journalists are forced to contend with the additional burden of verifying content authenticity while the public grows increasingly skeptical of media reports{\cite{wahl2021conjecturing}}. This skepticism can lead to a `cry wolf' scenario, where even legitimate news {contents are} doubted, contributing to a disconcerting post-truth era where facts are fungible, and {the t}ruth is subjective.

\subsection{Erosion of Public Trust}

The cumulative effects of unchecked deepfakes and misinformation are the erosion of public trust{\cite{pawelec2022deepfakes}}. When netizens cannot trust their eyes or ears, they can become cynical and disengaged. This disengagement poses risks not just to political processes but to the social fabric that binds communities together. Without trust, conspiracy theories flourish, scientific consensus is questioned, and social polarization deepens.

\subsection{Legal and Ethical Dilemmas}

The rise of AIGC has also precipitated legal and ethical dilemmas{\cite{ohman2020introducing}}. Current laws are ill-equipped to handle the nuances of deepfakes, often lagging behind technological advancements. Ethically, the implications are just as complex as {creating and distributing deepfakes of people without their consent, violating their rights.}

\section{Technical Defense Mechanisms}
{This section discusses the technological, strategic, and policy-oriented defense approaches that can mitigate the risks associated with AIGC.}
Since the realistic construction of deepfakes and dissemination of mis/disinformation {have} become more sophisticated with the advancement of {LM-based GenAI tools}, developing robust technical defense mechanisms is {a complex agenda. Below, we} outline current and emerging technologies aimed at detecting and countering AIGC {and} the challenges inherent in their deployment.

\subsection{Detection Algorithms}

Detection is the first line of defense against AI-generated false content. Algorithms designed to identify deepfakes typically analyze various data points that may indicate manipulation, such as inconsistencies in lighting, unnatural blinking patterns, or irregularities in skin texture. Advances in machine learning have led to the development of models that can scrutinize video frames for signs of alteration at a pixel level, often with the aid of deep learning techniques similar to those used to create deepfakes{\cite{mirsky2021creation}}. Audio deepfake detection similarly analyzes vocal patterns, looking for subtle signs of manipulation that may not be apparent to the human ear. These include irregularities in speech patterns, breathing sounds, and background noises{\cite{mirsky2021creation}}. The challenge lies in the fact that as detection algorithms become more sophisticated, so too do the methods for creating deepfakes, leading to an ongoing arms race between creators and detectors.

\subsection{AI-Driven Authentication Methods}

In addition to detection, authentication methods aim to verify the origin and integrity of content. Digital watermarking, for instance, involves embedding a hidden and unique pattern or code within the content at the time of creation, which can later be used to confirm its authenticity{\cite{wang2023synthetic}}. Blockchain technology offers another layer of security by providing a decentralized and immutable ledger of content creation and distribution, making unauthorized alterations easily traceable. Another approach is the use of biometric authentication, which employs unique biological characteristics such as facial recognition patterns, voiceprints, or even typing rhythms to confirm the identity of individuals in digital media{\cite{campbell2022preparing}}. These methods, however, must balance the need for security with concerns about privacy and the potential for misuse.

\subsection{Machine Learning-Based Authentication Techniques}

Machine learning is not only used to create deepfakes but can also be harnessed to combat them. Models can be trained to recognize the digital `fingerprints' left by the AI models that generate deepfakes. These fingerprints are often subtle flaws or patterns in the generated content that are consistent with the training data or generation method used{\cite{mitra2021machine}}. By analyzing these fingerprints, machine learning algorithms can identify whether content has been artificially generated or altered.

\subsection{Limitations and Challenges of Current Technologies}

While these technologies show promise, they are not without limitations. Deepfake creation techniques are evolving rapidly, and detection methods must continually adapt to keep pace~\cite{zhao2021multi}. Moreover, the computational resources required to analyze large volumes of content in real time are substantial, and false positives remain a concern. Another challenge is the ease of access to deepfake generation tools, which can be used by individuals with minimal technical expertise, further complicating detection efforts{\cite{masood2023deepfakes}}. Additionally, the adaptability of AI means that as soon as a detection method becomes effective, new techniques are developed to circumvent it. This cat-and-mouse dynamic requires a proactive and dynamic approach to defense mechanism development.

\subsection{The Need for Open Collaboration}

Given the scale and complexity of the challenge, open collaboration between academia, industry, and government is necessary. Sharing data, research findings, and strategies can accelerate the development of effective defense mechanisms{\cite{krishna2020deepfakes}}. Transparency in the functioning of detection and authentication technologies is also crucial to build trust and ensure these tools are used responsibly.

\section{Cross-Platform Strategies}

The digital ecosystem's interconnected nature necessitates cross-platform strategies to combat the spread of deepfakes and mis/disinformation effectively. This section outlines a collaborative approach that spans various stakeholders, including social media companies, technology firms, content creators, and end-users.

\subsection{The Role of Social Media and Technology Companies}

Social media platforms are the primary battlegrounds for the spread of deepfakes and mis/disinformation due to their vast reach and the speed at which content can go viral. These companies have a responsibility to actively monitor and mitigate the spread of fake content. Strategies include {\cite{hancock2021social}}:
\begin{itemize}
\item Content Moderation Enhancements: Using a combination of AI-driven and human moderation to detect and flag deepfakes.
\item Partnerships with Fact-Checkers: Collaborating with independent fact-checking organizations to verify content.
\item User Reporting Mechanisms: Empowering users to report suspicious content, which can then be reviewed by specialized teams.
\item Transparency Reports: Publishing regular reports on the number of deepfakes detected and the actions taken.
\item User Education: Providing educational resources to help users spot and understand the nature of deepfakes. 
\end{itemize}

\subsection{Collaborative Filtering and Fact-Checking Initiatives}

Collaborative filtering involves leveraging the collective effort of platform users to identify and filter out disinformation{\cite{unver2023emerging}}. This can be facilitated through:
\begin{itemize}
\item Community-Driven Moderation: Enabling community moderators to review and moderate content within their domains of expertise.
\item Crowdsourced Verification: Utilizing crowdsourcing to gather user input on the authenticity of content.
\item Real-Time Fact-Checking: Implementing systems that provide live fact-checking during events, speeches, and debates. 
\end{itemize}

\subsection{User-Centric Approaches}

Putting users at the center of the defense strategy involves education and empowerment {\cite{gupta2020eyes}}. This includes:
\begin{itemize}
\item Digital Literacy Programs: Educating the public on digital media, the existence of deepfakes, and the importance of critical thinking online.
\item Critical Media Literacy: Encouraging users to question the source and intent behind the content they consume.
\item Promotion of Verified Content: Boosting the visibility of content from verified and reputable sources. 
\end{itemize}

\subsection{Community Guidelines and Enforcement}

Platforms must establish clear community guidelines that define acceptable use and the consequences of spreading deepfakes and mis/disinformation {\cite{kikerpill2021dealing}}. Enforcement actions may include:
\begin{itemize}
\item Content Removal: Removing or demoting content that violates platform policies.
\item Account Suspension: Temporarily or permanently suspending accounts that repeatedly disseminate fake content.
\item User Feedback: Informing users when they have interacted with or shared false content. 
\end{itemize}

\subsection{Developing Standardized Protocols}

To streamline cross-platform efforts, there is a need for standardized protocols for content verification, data sharing, and incident response. This could involve {\cite{shiohara2022detecting}}:
\begin{itemize}
\item Interoperable Verification Tags: Creating tags that indicate content has been verified, which can be recognized across different platforms.
\item Data Sharing Agreements: Establishing agreements to share data on deepfakes and misinformation trends and techniques.
\item Joint Response Frameworks: Developing coordinated response plans for widespread disinformation campaigns. 
\end{itemize}

\section{Ethical Considerations}

The ethical implications of deepfakes and misinformation are as vast and complex as their technical and social counterparts~\cite{diakopoulos2021anticipating}. This section explores the moral landscape that AIGC presents, the responsibilities of creators and disseminators, and the overarching need for ethical guidelines to shape the evolution of AI technologies.

\subsection{Ethical AI Development and Use}

The development of AI technologies is not value-neutral; it reflects the biases, priorities, and ethical orientations of its creators. Therefore, the following needs to be addressed.
\begin{itemize}
\item Bias and Fairness: There is a need for ethical AI development that actively seeks to minimize biases in training data and algorithms, ensuring fairness and non-discrimination~\cite{giovanola2023beyond}.
\item Transparency: AI systems should be developed with transparency in mind, allowing for traceability and explainability in the AI's decision-making processes~\cite{ehsan2021expanding}.
\item Accountability: Developers and users of AI must be accountable for the outcomes of their technologies, particularly when they impact public opinion or infringe on personal rights~\cite{henriksen2021situated}.
\end{itemize}

\subsection{The Balance between Innovation and Regulation}

There is a delicate balance to be maintained between encouraging innovation in AI and implementing regulations that protect against its misuse:
\begin{itemize}
\item Innovation-Friendly Policies: Policies should aim to foster innovation and the beneficial applications of AI while guarding against risks.
\item Proactive Ethical Design: AI should be designed proactively with ethical considerations in mind, rather than retroactively applying ethical standards to existing technologies.
\end{itemize}

\subsection{Future Outlook and Philosophical Implications}

AI's capabilities force us to confront deep philosophical questions about the nature of truth, reality, and human experience:
\begin{itemize}
\item Ontological Questions: As AI blurs the lines between reality and simulation, we must address the ontological status of experiences and entities created by AI.
\item Epistemological Considerations: The proliferation of deepfakes calls into question the basis of knowledge and the conditions under which we can claim to know something as true or false.
\item Human Agency and Autonomy: There is a need to consider how AI impacts human agency and autonomy, particularly when individuals are subject to AI-generated representations without their consent.
\end{itemize}

\subsection{The Ethical Use of Deepfakes}

While deepfakes are often discussed in negative terms, they also have potentially positive applications:
\begin{itemize}
\item Artistic and Educational Uses: Deepfakes can be used for legitimate artistic expression or educational purposes, such as recreating historical speeches~\cite{danry2022ai}.
\item Medical and Therapeutic Applications: There are possibilities for using deepfake technology in medical simulations or therapeutic settings~\cite{yang2022can}.
\end{itemize}

\section{Proposed Integrated Defense Framework}

The multifaceted nature of the threats posed by deepfakes and mis/disinformation necessitates a comprehensive response~\cite{kelly2024s}. This section proposes an integrated defense framework that synthesizes technological, strategic, policy-oriented, and educational responses to these threats.

\subsection{Design of the Integrated Defense Framework}

The proposed framework is designed with four key pillars:
\begin{itemize}
\item Technological Solutions: Incorporating advanced detection algorithms, AI-driven authentication methods, and machine learning-based authentication techniques.
\item Strategic Initiatives: Implementing cross-platform strategies, including content moderation enhancements and collaborative filtering.
\item Policy and Regulation: Developing new legislation and ethical guidelines that clearly define and impose penalties for the creation and distribution of deepfakes.
\item Education and Public Awareness: Launching comprehensive educational programs and public awareness campaigns to improve media literacy and critical thinking.
\end{itemize}

\subsection{Implementation of the Framework}

For effective implementation, the framework requires:
\begin{itemize}
\item Multi-Stakeholder Collaboration: Coordination among governments, tech companies, academia, and civil society to ensure a united front against deepfakes.
\item Resource Allocation: Commitment of financial, human, and technological resources to support the framework's initiatives.
\item Adaptive Strategies: Continuous adaptation of strategies to address the evolving nature of deepfake and misinformation tactics~\cite{jaeger2021arsenals}.
\end{itemize}

\subsection{Case Study: Applying the Framework in a Simulated Environment}

To validate the framework, a simulated environment that replicates the complex ecosystem of media platforms and AIGC can be created~\cite{liu2023blockchain}. Here, the framework's components would be tested against various attack scenarios to assess their effectiveness and identify areas for improvement.

\subsection{Analysis of Framework Effectiveness}

Evaluating the effectiveness of the defense framework involves:
\begin{itemize}
\item Monitoring and Evaluation: Regular assessment of each pillar's performance in detecting and countering deepfakes.
\item Feedback Mechanisms: Systems for collecting feedback from stakeholders to inform the iterative improvement of the framework.
\item Benchmarking: Setting benchmarks for success and conducting comparative analysis with other defense strategies.
\end{itemize}

\subsection{Potential Unforeseen Consequences and Mitigation Strategies}

While the framework aims to be comprehensive, there may be unforeseen consequences, such as over-censorship or the stifling of innovation. Mitigation strategies include:
\begin{itemize}
\item Ethical Oversight: Establishing ethical oversight committees to review the impact of defense measures.
\item Balanced Approach: Ensuring a balanced approach that respects freedom of expression while protecting against misinformation.
\item Rapid Response Protocols: Developing protocols for rapidly addressing negative consequences as they arise.
\end{itemize}

\section{Discussion}

The emergence of deepfakes and the proliferation of mis/disinformation through the use of advanced AI models pose a significant threat to the integrity of information, necessitating a multi-pronged approach to mitigation~\cite{langmia2023black}. This discussion evaluates the proposed solutions, explores potential unintended consequences, and highlights ongoing challenges and areas for future research.

\subsection{Analysis of the Proposed Solutions' Effectiveness}

The proposed integrated framework's effectiveness hinges on the synergy between its components:
\begin{itemize}
\item Technological Efficacy: The rapid detection of deepfakes is crucial. However, as the technology to create deepfakes becomes more sophisticated, detection methods may need to become more specialized, potentially leading to an arms race between creation and detection capabilities.
\item Strategic Resilience: Cross-platform strategies emphasize the need for a coordinated response to misinformation. The scalability of such initiatives is vital, as is the ability to adapt quickly to new forms of disinformation.
\item Policy Impact: The effectiveness of policy measures will largely depend on their enforcement and the international community's willingness to adopt and implement harmonized standards.
\item Educational Outcomes: Long-term, the success of educational programs in enhancing the public's ability to discern true from false information may be one of the most sustainable defenses against misinformation.
\end{itemize}

\subsection{Open Challenges and Areas for Future Research}

Several challenges remain open, requiring ongoing attention:
\begin{itemize}
\item Technological Advancement: Keeping defensive measures up-to-date with the latest advancements in AI and deepfake technologies.
\item Global Cooperation: Achieving a consensus on international standards and cooperation in the face of geopolitical tensions and differing national interests.
\item Public Engagement: Ensuring continued public engagement and understanding in the face of ``fatigue'' around the topic of misinformation.
Future research areas are plentiful, including:
\item Behavioral Insights: Gaining a deeper understanding of why people create and spread misinformation, and how they are influenced by it.
\item Economic Models: Developing economic models to understand the incentives behind the spread of deepfakes and misinformation~\cite{kshetri2023economics}.
\item Technological Innovations: Exploring new technological innovations that can preemptively address the creation of deepfakes.
\end{itemize}

\section{Conclusion}

%{The paper emphasizes the critical threat posed by deepfakes and generative AI to the global information ecosystem, highlighting the need for a multifaceted defense. While technological solutions are crucial, they face ongoing challenges from evolving AI falsification techniques. Platforms must strategically prioritize information integrity without compromising freedom of expression, and international policy frameworks need to be adaptable in the borderless digital landscape. The cornerstone of long-term resilience lies in public education and awareness initiatives. The paper advocates for a collaborative, proactive approach involving technological development, media literacy education, and regulatory strategies. The battle against deepfakes is portrayed not merely as a technical challenge but as a societal one that necessitates a collective commitment to truth and reality. The suggested strategies provide a comprehensive defense framework but underscore the need for continuous vigilance, innovation, and cooperation across stakeholders. Ultimately, the preservation of factual discourse and democratic structures relies on concerted efforts from technologists, educators, policymakers, and an informed citizenry.}
The paper emphasizes the critical role of frontier AI in countering the profound threat of deepfakes and generative AI to global information ecosystems. It underscores the need for a comprehensive, multi-faceted defense strategy that evolves in tandem with frontier AI advancements. The paper highlights the importance of developing sophisticated technological solutions, adaptable international policies, and enhancing public education in media literacy to effectively combat these threats. Advocating for a collaborative approach, it integrates the advancements in frontier AI with regulatory strategies and media literacy education, framing the battle against deepfakes as not only a technical challenge but a broader societal issue.

\section*{Acknowledgement}

This research is partly supported by the Singapore Ministry of Education Academic Research Fund under Grant Tier 1 RG90/22, Grant Tier 1 RG97/20, Grant Tier 1 RG24/20 and Grant Tier 2 MOE2019-T2-1-176; and partly by the Nanyang Technological University (NTU)-Wallenberg AI, Autonomous Systems and Software Program (WASP) Joint Project.

\bibliographystyle{IEEEtran}
\bibliography{related}

\end{document}